\documentclass[journal]{IEEEtran}
\usepackage{cite}
\usepackage{subfig}
\usepackage{amsmath,amssymb,amsfonts,amsthm}
\usepackage{algorithmic}
\usepackage{graphicx}
\usepackage{hyperref}
\usepackage{multirow}
\usepackage{multicol}
\usepackage{booktabs}
\usepackage{makecell}
\usepackage{tabularx}
\usepackage{gensymb}
\hypersetup{colorlinks=true,urlcolor=red}
\usepackage{bm} 
\usepackage{xspace}

\newcommand{\xmath}[1] {\ensuremath{#1}\xspace}
\newcommand{\blmath}[1] {\xmath{\bm{#1}}}

\newcommand{\G}{\blmath{G}}

\newcommand{\wb}{{\blmath w}}
\newcommand{\xb}{{\blmath x}}
\newcommand{\yb}{{\blmath y}}

\newcommand{\Hc}{\mathcal{H}}

\newcommand{\Cc}{\mathcal{C}}

\newcommand{\Sc}{\mathcal{S}}

\newcommand{\Xc}{\mathcal{X}}
\newcommand{\Yc}{\mathcal{Y}}

\newcommand{\Pc}{{{\mathcal P}}}

\newcommand{\Kd}{\mathbb{K}}

\newcommand{\beq}{\begin{equation}}
\newcommand{\eeq}{\end{equation}}
\newcommand{\beqa}{\begin{eqnarray}}
\newcommand{\eeqa}{\end{eqnarray}}

\begin{document}

\title{Unpaired Deep Learning for Pharmacokinetic Parameter Estimation from Dynamic Contrast-Enhanced MRI}
\date{\vspace{-4ex}}

\author{Gyutaek Oh, 
		Won-Jin Moon, 
		and~Jong~Chul~Ye,~\IEEEmembership{Fellow,~IEEE}
\thanks{G. Oh is with the Department of Bio and Brain Engineering, 
		Korea Advanced Institute of Science and Technology (KAIST), 
		Daejeon 34141, Republic of Korea (e-mail: okt0711@kaist.ac.kr).
		J. C. Ye is with Kim Jaechul Graduate School of AI,
		Korea Advanced Institute of Science and Technology (KAIST), 
		Daejeon 34141, Republic of Korea (e-mail: jong.ye@kaist.ac.kr).
		W. Moon is with Department of Radiology, Konkuk University School of Medicine, Seoul 05030, Republic of Korea (e-mail: mdmoonwj@gmail.com).}
\thanks{W.J. Moon and J.C. Ye are corresponding authors.} 
}

\maketitle

\begin{abstract}
DCE-MRI provides information about vascular permeability and tissue perfusion through the acquisition of pharmacokinetic parameters.
However, traditional methods for estimating these pharmacokinetic parameters involve fitting tracer kinetic models, which often suffer from computational complexity and low accuracy due to noisy arterial input function (AIF) measurements.
Although some deep learning approaches have been proposed to tackle these challenges, most existing methods rely on supervised learning that requires paired input DCE-MRI and labeled pharmacokinetic parameter maps.
This dependency on labeled data introduces significant time and resource constraints, as well as potential noise in the labels, making supervised learning methods often impractical.
To address these limitations, here we present a novel unpaired deep learning method for estimating both pharmacokinetic parameters and the AIF using a physics-driven CycleGAN approach.
Our proposed CycleGAN framework is designed based on the underlying physics model, resulting in a simpler architecture with a single generator and discriminator pair.
Crucially, our experimental results indicate that our method, which does not necessitate separate AIF measurements, produces more reliable pharmacokinetic parameters than other techniques.
\end{abstract}


\IEEEpeerreviewmaketitle

\section{Introduction}\label{sec:introduction}
\IEEEPARstart{D}{ynamic} contrast-enhanced magnetic resonance imaging (DCE-MRI) is an invaluable T1-weighted imaging technique utilized in the diagnosis and management of various diseases \cite{verma2012overview,armitage2011use,van2016blood}.
In DCE-MRI, a contrast agent (CA) such as gadobutrol (Gd-BT-DO3A) is injected, and the changes in the CA concentration within different tissues are captured through a series of T1-weighted images over time.
This CA concentration data enables the estimation of pharmacokinetic (PK) parameters associated with vascular permeability and tissue perfusion \cite{sourbron2013classic}.
These PK parameters offer valuable insights into the physiological characteristics of tissues.

There exist mathematical relationships between the concentration of CA and PK parameters, which are captured by tracer kinetic (TK) models such as extended Tofts (eTofts) model \cite{tofts1999estimating}, Patlak model \cite{patlak1983graphical}, etc.
Consequently, the estimation of PK parameters involves fitting one of these TK models to the time series of CA concentration for each voxel.
The common approach for model fitting is the nonlinear least squares (NLLS) method \cite{branch1999subspace}, but it is computationally intensive due to a large number of iterations.
More importantly, NLLS methods depend on inputs like the arterial input function (AIF), which, if obtained from inappropriate voxels or affected by high noise, can significantly compromise accuracy.
Another method, linear least squares (LLS) \cite{flouri2016fitting}, was proposed to reduce computational complexity and runtime, but it exhibits decreased accuracy when noise levels are high.

Recently, deep learning methods have emerged for PK parameter estimation from DCE-MRI \cite{choi2020improving,ulas2019convolutional,fang2021convolutional}.
These methods offer reduced runtime and yield estimation results comparable to NLLS or LLS methods.
However, existing deep learning methods rely on supervised learning, necessitating matched pairs of input DCE-MRI and label PK parameter maps.
Obtaining these label PK parameter maps traditionally entails time-consuming NLLS or LLS estimation.
Furthermore, if the label data contains high noise levels, especially due to the incorrect AIF estimation, the performance of supervised methods may be adversely affected.

To address the limitations of conventional and supervised methods, here we propose an unpaired deep learning approach for directly estimating PK parameters from DCE-MRI.
Our method draws inspiration from the optimal transport-driven CycleGAN (OT-CycleGAN) \cite{sim2020optimal} which is applicable in scenarios where matched input-label pairs are unavailable.
Leveraging the known TK models relating PK parameters to DCE-MR images, our OT-CycleGAN employs a single generator and discriminator, resulting in reduced computational complexity.
Another important contribution is that our model does not rely on AIF for PK parameter estimation as it allows for AIF extraction.
Experimental results demonstrate that our proposed method, which does not necessitate separate AIF measurements, achieves competitive performance compared to supervised learning, despite being trained with an unpaired dataset.

This paper is organized as follows.
First, we present a comprehensive review of the TK models and optimal transport-driven CycleGAN in Section \ref{sec:backgrounds}.
The theoretical foundations of our proposed method are elaborated upon in Section \ref{sec:theory}.
Following this, we outline the experimental methods (Section \ref{sec:method}) and present the corresponding results (Section \ref{sec:result}).
In Section \ref{sec:discussion}, we engage in a through discussion of our findings, and finally, in Section \ref{sec:conclusion}, we provide a conclusive summary of our paper.

\section{Backgrounds}\label{sec:backgrounds}
\subsection{Tracer Kinetic Models}
Tracer kinetic (TK) models \cite{sourbron2013classic,heye2016tracer} establish the connections between the concentration of the contrast agent (CA) and the pharmacokinetic (PK) parameters.
In this study, we will examine two commonly employed TK models in DCE-MRI: the extended Tofts model and the Patlak model.

\subsubsection{Extended Tofts Model}
\begin{figure}[!t]
\centerline{\includegraphics[width=0.9\columnwidth]{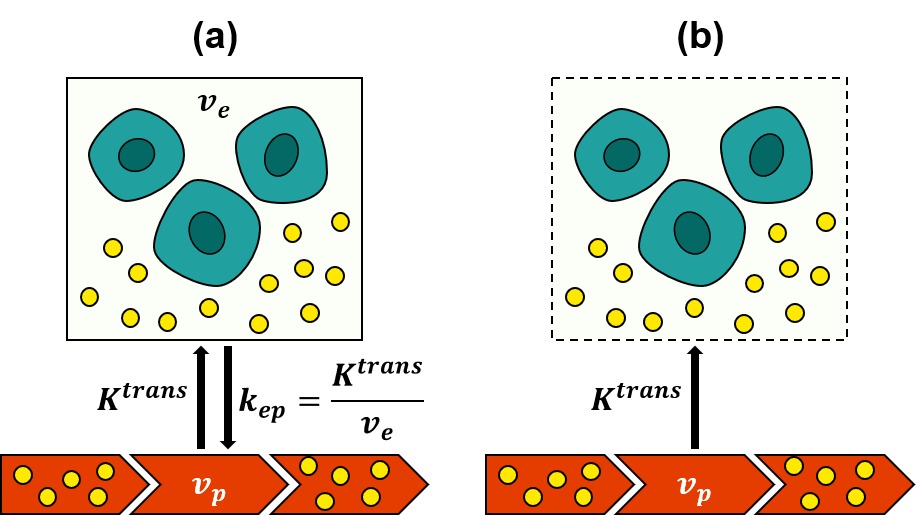}}
\caption{Two commonly utilized tracer kinetic models in DCE-MRI: (a) the extended Tofts model, and (b) the Patlak model.
The parameters in the models are as follows: $K^{trans}$ represents the volume transfer constant, $v_p$ denotes the fractional plasma volume, and $v_e$ signifies the fractional interstitial volume.
}
\label{fig:tk_model}
\end{figure}

The extended Tofts (eTofts) model \cite{tofts1999estimating} is a TK model that assumes the highly perfused tissue condition (Fig.~\ref{fig:tk_model}(a)).
Specifically, the eTofts model assumes the presence of infinite plasma flow.
According to the eTofts model, the CA concentration in the tissue, denoted as $C_t(t)$, is represented by the following equation:
\begin{equation}\label{eq:etofts}
    C_t(t) = v_pC_p(t) + K^{trans}\int_0^t C_p(\tau)e^{-\frac{K^{trans}}{v_e}(t-\tau)} d\tau
\end{equation}
where $C_p(t)$ represents the CA concentration in the plasma, $v_p$ corresponds to the fractional plasma volume, $v_e$ denotes the fractional interstitial (or extravascular extracellular space) volume, and $K^{trans}$ denotes the volume transfer constant from the plasma to the interstitium.
Consequently, the estimation of the eTofts model involves determining three PK parameters: $K^{trans}$, $v_p$, and $v_e$.

\subsubsection{Patlak Model}
When the backflux from the interstitium to the plasma is negligible, as depicted in Fig. \ref{fig:tk_model}(b), the eTofts model can be simplified to a two-parameter model, expressed as follows:
\begin{equation}\label{eq:patlak}
    C_t(t) = v_pC_p(t) + K^{trans}\int_0^t C_p(\tau) d\tau.
\end{equation}
Hence, this two-parameter model is known as the Patlak model \cite{patlak1983graphical}, which can be considered a special case of the eTofts model.
In the Patlak model, $K^{trans}$ and $v_p$ are estimated.

\subsubsection{Contrast Agent Concentration to DCE-MRI}
If the time series DCE-MRI signal, denoted as $S(t)$, is acquired using the spoiled gradient echo, it can be converted to the corresponding CA concentration, denoted as $C_t(t)$, by \cite{di2009accuracy,chao2017tumourmetrics}
\begin{equation}\label{eq:s_to_c}
    \begin{split}
    &C_t(t)\\
    &= -\frac{1}{B}\left[A+\ln\left(\frac{\frac{S(t)}{S_0}\frac{1-e^{-A}}{1-\cos(\alpha)e^{-A}} - 1}{\frac{S(t)}{S_0}\cos(\alpha)\frac{1-e^{-A}}{1-\cos(\alpha)e^{-A}} - 1}\right)\right]
    \end{split}
\end{equation}
where $S_0$ is the pre-contrast image intensity, $\alpha$ is the flip angle, and
\begin{equation}\notag
    A = \frac{TR}{T_1}, \quad B = r_1TR
\end{equation}
where $TR$ represents the repetition time, $T_1$ signifies the $T_1$ relaxation, and $r_1$ denotes the relaxivity of the contrast agent.
Conversely, if we intend to obtain $S(t)$ from the CA concentration $C_t(t)$, derived from either the eTofts or Patlak model, Eq. \eqref{eq:s_to_c} can be transformed as follows:
\begin{equation}\label{eq:c_to_s}
    S(t) = \frac{(1 - e^{-A-B\cdot C_t(t)})(1 - \cos(\alpha)e^{-A})}{(1 - \cos(\alpha)e^{-A-B\cdot C_t(t)})(1 - e^{-A})}S_0.
\end{equation}

\subsection{Deep Learning Methods for Pharmacokinetic Parameter Estimation}
Due to the high computational complexity and additional requirements such as the arterial input function (AIF), deep learning methods have emerged as an alternative for PK parameter estimation.
Choi et al. \cite{choi2020improving} introduced conditional generative adversarial networks to translate AIF from DCE-MRI to AIF from dynamic susceptibility-enhanced MRI (DSC-MRI).
By leveraging the higher reliability of AIF derived from T2*-weighted DSC-MRI, compared to the low signal intensity T1-weighted DCE-MRI, they demonstrated improved reliability of estimated PK parameters.
However, this method does not directly estimate the PK parameters through deep learning.

Ulas et al. \cite{ulas2019convolutional} and Fang et al. \cite{fang2021convolutional} proposed the use of convolutional neural networks (CNNs) for the direct estimation of PK parameters.
Their approaches employed CNNs with both local and global pathways, incorporating forward physical loss.
These methods exhibited reliable results and reduced runtime compared to traditional NLLS or LLS methods.
However, both approaches rely on supervised learning, necessitating paired input DCE-MRI images and corresponding labeled PK parameter maps.
The acquisition of these labeled PK parameter maps for supervised training typically requires time-consuming NLLS or LLS estimations.
Moreover, the accuracy of the label PK parameter maps heavily depends on the accuracy of AIF measurement.
As a result, the process suffers from extended estimation time or degraded data quality.

To overcome these challenges, we proposed an unpaired deep learning method for the direct estimation of PK parameters using a physics-driven CycleGAN.
This approach provides solutions to the challenges linked with obtaining labeled training data, facilitating direct approximation without requiring paired DCE-MRI images, matched PK parameter maps, and AIF measurements.

\subsection{Optimal Transport Driven CycleGAN}
In our previous investigations \cite{sim2020optimal,lim2020cyclegan}, we uncovered the connection between the architecture of CycleGAN and the dual formulation of an optimal transport problem.
In this section, we present a concise overview of the theoretical foundations explored in those studies.

Consider the following measurement model:
\begin{equation}
    \yb = \Hc \xb + \wb
\end{equation}
where $\xb \in \Xc$ and $\yb \in \Yc$ denote the unknown image and the noisy measurement, respectively, $\wb$ is the measurement noise, and $\Hc : \Xc \mapsto \Yc$ is the known deterministic imaging operator.
In supervised learning, the objective is to learn the relationship between image $\xb$ and its corresponding measurement $\yb$ pairs.
However, in unpaired learning scenarios, there is no availability of matched image-measurement pairs.
Consequently, the goal shifts towards matching the probability distributions instead of establishing direct correspondences between individual samples from the unpaired sets of images and measurements.
This is accomplished by identifying transportation maps that facilitate the transportation of probability distributions between the two spaces.

\begin{figure*}[!t]
\centering
\includegraphics[width=0.7\linewidth]{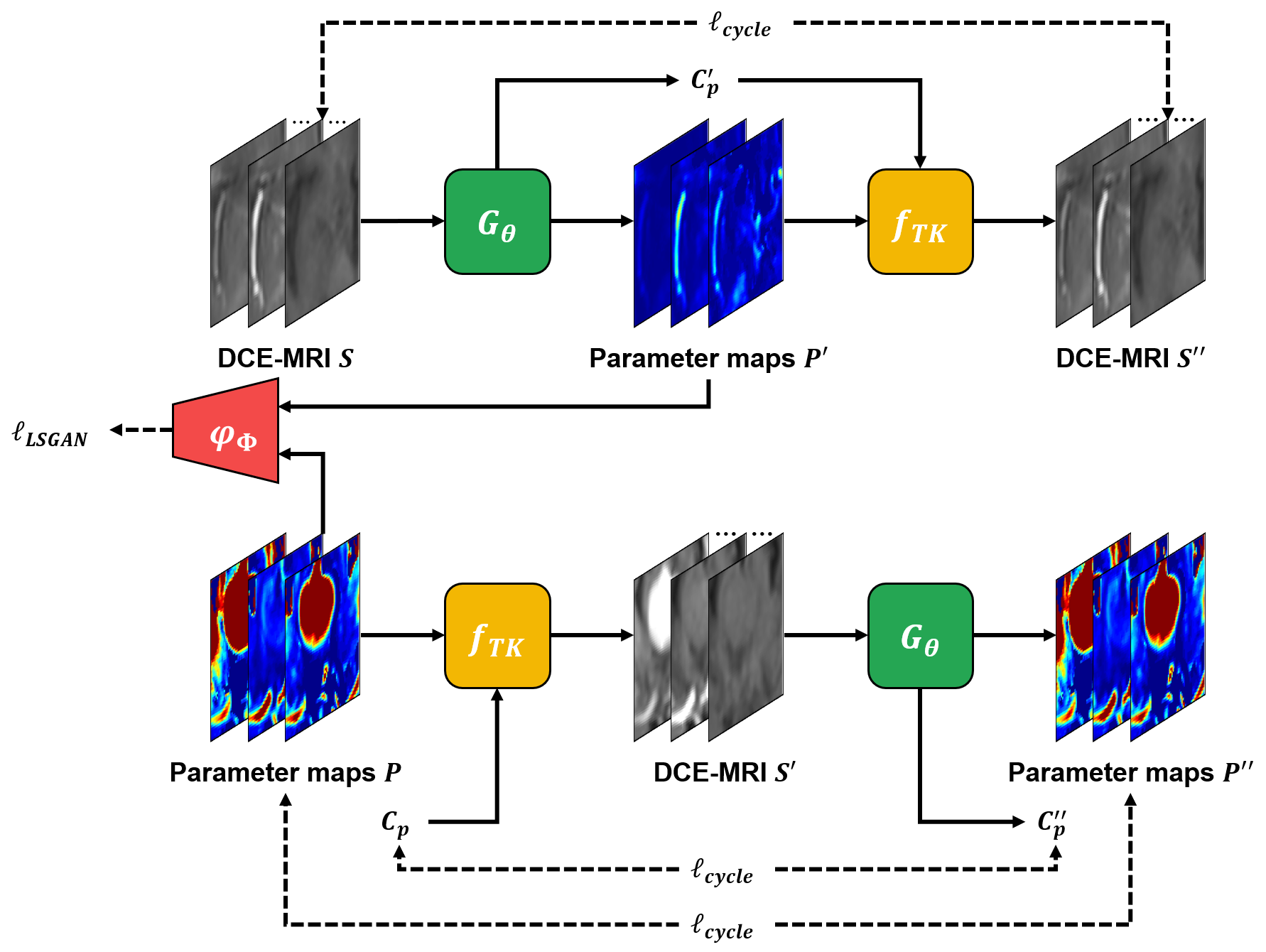}
\caption{The architecture of the proposed CycleGAN for PK parameter estimation consists of key components:
$\G_\theta$ representing the generator, $f_{TK}$ denoting the known tracer kinetic model, $\varphi_\Phi$ representing the discriminator, and $C_p$ representing the arterial plasma concentration.
The proposed CycleGAN utilizes a single generator and a single discriminator due to the deterministic nature of physics model mapping PK parameters to DCE-MRI.
}
\label{fig:cycleDCE}
\end{figure*}

Let us consider the target image space $\Xc$ and the measurement space $\Yc$, which are associated with probability measures $\mu$ and $\nu$, respectively.
To facilitate the transportation of mass from $(\Xc, \mu)$ to $(\Yc, \nu)$, we employ the forward operator $\Hc$.
Conversely, the generator $G_\Theta : \Yc \mapsto \Xc$, parameterized by $\Theta$, performs the mass transport from $(\Yc, \nu)$ to $(\Xc, \mu)$.
In this context, the following transportation cost was proposed for the optimal transport problem \cite{sim2020optimal}:
\begin{equation}\label{eq:tcost}
    c(\xb, \yb; \Theta) := \|\yb - \Hc \xb\| + \|G_\Theta(\yb) - \xb\|,
\end{equation}
which denotes the sum of the distance between a training sample and a transported sample in each space.
The objective of the optimal transport is not limited to minimizing the sample-wise cost computed using Eq. \eqref{eq:tcost}.
Instead, its goal is to minimize the average trasnport cost.
Specifically, the optimal transport problem aims to identify the joint distribution $\pi$ that results in the minimum average transport cost.
Thus, the problem is formulated as finding the $\pi$ that minimizes the average transport cost:
\begin{equation}\label{eq:unpaired}
    \inf_{\pi \in \Pi(\mu, \nu)} \int_{\Xc \times \Yc} c(\xb, \yb; \Theta) d\pi(\xb, \yb),
\end{equation}
where $\Pi(\mu, \nu)$ denotes the set of joint measures whose marginal distributions in $\Xc$ and $\Yc$ are $\mu$ and $\nu$, respectively.

Using the transportation cost in \eqref{eq:tcost}, the Kantorovich dual formulation \cite{villani2008optimal} is given by \cite{sim2020optimal}
\begin{equation}\label{eq:minmax}
    \min_\Theta \Kd(\Theta, \Hc) = \min_\Theta \max_\Phi \ell(\Theta, \Phi),
\end{equation}
where
\begin{equation}
    \ell(\Theta; \Phi) = \gamma \ell_{cycle}(\Theta) + \ell_{WGAN}(\Theta; \Phi).
\end{equation}
Here, $\gamma$ is a suitable hyperparameter, $\ell_{cycle}$ is the cycle-consistency loss, and $\ell_{WGAN}$ is the Wasserstein GAN (WGAN) loss \cite{martin2017wasserstein}.
More specifically, $\ell_{cycle}$ and $\ell_{WGAN}$ are given by
\begin{equation}\label{eq:cycle}
    \begin{split}
    \ell_{cycle}(\Theta) = \int_\Xc \|\xb - G_\Theta(\Hc \xb)\| d\mu(\xb)\\
    + \int_\Yc \|\yb - \Hc G_\Theta(\yb)\| d\nu(\yb),
    \end{split}
\end{equation}
\begin{equation}\label{eq:WGAN}
    \begin{split}
    &\ell_{WGAN}(\Theta; \Phi)\\
    &= \int_\Xc \varphi_\Phi (\xb) d\mu(\xb) - \int_\Yc \varphi_\Phi(G_\Theta(\yb)) d\nu(\yb).
    \end{split}
\end{equation}
It is worth noting that in the proposed architecture, there is only one discriminator, denoted as $\varphi_\Phi$, due to the known nature of the forward operator $\Hc$.
As a result, there is no need for the discriminator to compete with the forward operator.
This simplicity in the CycleGAN architecture will be further elucidated in the subsequent explanation.

\section{Theory}\label{sec:theory}
Now, we will employ the optimal transport-driven CycleGAN for PK parameter estimation.
In our specific scenario, a time series DCE-MRI signal $S\in\Sc$ and a PK parameter map $P\in\Pc$ correspond to a measurement and an unobserved image, respectively.
To be more precise, we can formulate the forward model mapping PK parameter maps to DCE-MRI as follows:
\begin{equation}
    S = f_{TK}(P, C_p)
\end{equation}
where $f_{TK}$ is the function composed by the TK models (Eqs. \eqref{eq:etofts} or \eqref{eq:patlak}) and \eqref{eq:c_to_s}.
Here, the availability of the arterial plasma concentration $C_p\in\Cc$ is necessary, and it is feasible to utilize $C_p(t)$ derived from the measured AIF.

However, if the measured AIF is affected by noise or obtained from inappropriate voxels, it can significantly degrade the performance of our model.
To address this challenge, we propose a framework where the generator simultaneously estimates the PK parameter maps and the arterial plasma concentration $C_p$.
This is illustrated in Fig. \ref{fig:cycleDCE}.
Additionally, in the lower branch of the CycleGAN framework depicted in Fig. \ref{fig:cycleDCE}, we apply a cycle-consistency loss for $C_p$ to ensure that the estimated AIF remains consistent and avoids unusual estimations.

By identifying $P', C_p' = G_\Theta(S)$, $S' = f_{TK}(G_\Theta(S))$ and $P'', C_p'' = G_\Theta(f_{TK}(P, C_p))$, the cycle-consistency loss and WGAN loss for our OT-cycleGAN can be represented by
\begin{equation}\label{eq:cycle_pk}
    \begin{split}
    \ell_{cycle}(\Theta) &= \int_\Pc \|P - P''\| d\mu(P)\\
    &+ \int_\Sc \|S - S''\| d\nu(S)\\
    &+ \rho\int_\Cc \|C_p - C_p''\| d\kappa(C_p),
    \end{split}
\end{equation}
\begin{equation}
    \begin{split}
    &\ell_{WGAN}(\Theta; \Phi)\\
    &= \int_\Pc \varphi_\Phi (P) d\mu(P) - \int_\Sc \varphi_\Phi(P') d\nu(S),
    \end{split}
\end{equation}
where $\rho$ is a hyperparameter to control the weight of cycle-consistency loss, $\kappa$ denotes the probability
measure for $C_p$ distribution.

While the direct application of OT-CycleGAN can be utilized for PK parameter estimation, we choose to employ the least squares GAN (LSGAN) \cite{mao2017least} instead of WGAN to ensure more stable training.
The relationship between WGAN and LSGAN was previously discussed in previous work \cite{lim2020cyclegan}.
Consequently, the final cost function for OT-CycleGAN can be formulated as
\begin{equation}\label{eq:total}
    \ell(\Theta, \Phi) = \gamma\ell_{cycle}(\Theta) + \ell_{LSGAN}(\Theta;\Phi).
\end{equation}

\section{Method}\label{sec:method}
\subsection{Experimental Dataset}
In our experiments, we utilized two datasets acquired from Konkuk University Medical Center.
For both datasets, we employed gadobutrol as the contrast agent with a relaxivity value of $r_1=3.47$ mM$^{-1}$s$^{-1}$ \cite{szomolanyi2019comparison}.
A standard dose of gadobutrol (0.1 mmol/kg body weight; Bayer Healthcare) was administered with a 30 mL saline flush using an automatic injector, after the fourth dynamic scan, at a flow rate of 2 mL/s.
To account for T1 inhomogeneity during kinetic parameter calculations, T1 mapping was generated using a pre-contrast T1-weighted gradient echo series with six different flip angles (2 - 12$\degree$).

\subsubsection{DCE-MRI for Brain Tumor}
The first dataset, referred to as the `\textbf{Tumor Data}', consisted of DCE imaging data acquired from 116 patients with brain tumors.
An axial 3D DCE sequence was employed, comprising a dynamic series of 65 individual scans.
The imaging parameters for this sequence were as follows: TR = 2.80 ms, TE = 0.90 ms, flip angle = 10$\degree$, average = 1, field-of-view = 222 $\times$ 240 mm$^2$, slice thickness = 4 mm, acquisition matrix = 160 $\times$ 148, voxel size = 0.75 $\times$ 0.75 $\times$ 4.00 mm$^3$, acquisition time = 7 min, and time resolution = 6.5 s.
Among the patients, data from 98 individuals were utilized for training, while the remaining data were used for inference purposes.

\subsubsection{DCE-MRI for normal control and MCI workup}
The second dataset referred to as the `\textbf{MCI Data}', comprised imaging data obtained from subjects with normal cognition and mild cognitive impairment (MCI).
A coronal 3D DCE sequence was employed to acquire a dynamic series of 60 individual scans.
The imaging parameters for this sequence were as follows: TR = 3.72 ms, TE = 1.04 ms, filp angle = 10$\degree$, average = 1, field-of-view = 225 $\times$ 240 mm$^2$, slice thickness = 3 mm, acquisition matrix = 180 $\times$ 192, voxel size = 1.25 $\times$ 1.25 $\times$ 3.00 mm$^3$, acquisition time = 10 min, and time resolution = 10 s.
The acquisition time of 10 minutes was chosen considering both patient compliance and mathematical modeling requirements.
The coronal plane was selected for DCE imaging as it provides an optimal assessment of the hippocampi and temporal lobes.
For training purposes data from 196 subjects were utilized, while the evaluation of our method was performed using data from 36 subjects.

\subsubsection{DCE-MRI analysis}
Postprocessing and region of interest selection in the DCE imaging data were performed using NordicICE software (Version 4.1.3), with T1-volume imaging utilized for structural imaging.
The analysis was conducted by a trained researcher (Changmok Lee, with three years of experience) under the supervision of an expert neuroradiologist (Won-Jin Moon, with 20 years of neuroimaging expertise).
Both the researcher and supervisor were blinded to the clinical diagnosis.

For the conventional approaches, the arterial input function (AIF) was obtained using the semiautomatic method within the NordicICE software.
Following motion correction and noise modulation, automatic AIF search box detection was applied.
In the first dataset, the AIF search box was placed in the area of the internal carotid artery on the axial slice, while in the second dataset, it was placed in the area of the superior sagittal sinus on the coronal plane.
The final AIF curve was determined by calculating the mean of one to five different curves (up to 5 pixels) within the search box.
For each patient, three to five (or one to five in the second dataset) pixel-time curves were selected, exhibiting expected input function properties such as low first moment, high peak enhancement, and a satisfactory area under the curve.
After obtaining the AIF, we converted it to plasma concentration using the formula $C_p(t) = C_b(t) / (1 - \text{Hct})$, where $C_b(t)$ represents the measured AIF and Hct denotes the blood hematocrit.
Consistent with previous studies \cite{heye2016tracer,ulas2019convolutional}, we used a value of $\text{Hct} = 0.45$.

For the `Tumor Data', the eTofts permeability model was employed to estimate standard DCE perfusion parameters in the tissue \cite{padhani2002dynamic}.
Maps of $K^{trans}$, $v_p$, and $v_e$ derived from the eTofts model were calculated following AIF acquisition.
Conversely, for the `MCI Data', the Patlak model, considered optimal for low-leakage conditions \cite{heye2016tracer,thrippleton2019quantifying}, was utilized.
Specifically, the two parameters, $K^{trans}$ and $v_p$, were calculated using the Patlak model.
In the NordicICE software, the LLS method \cite{flouri2016fitting} was employed for estimating the PK parameters.

\begin{figure}[!t]
\centerline{\includegraphics[width=\columnwidth]{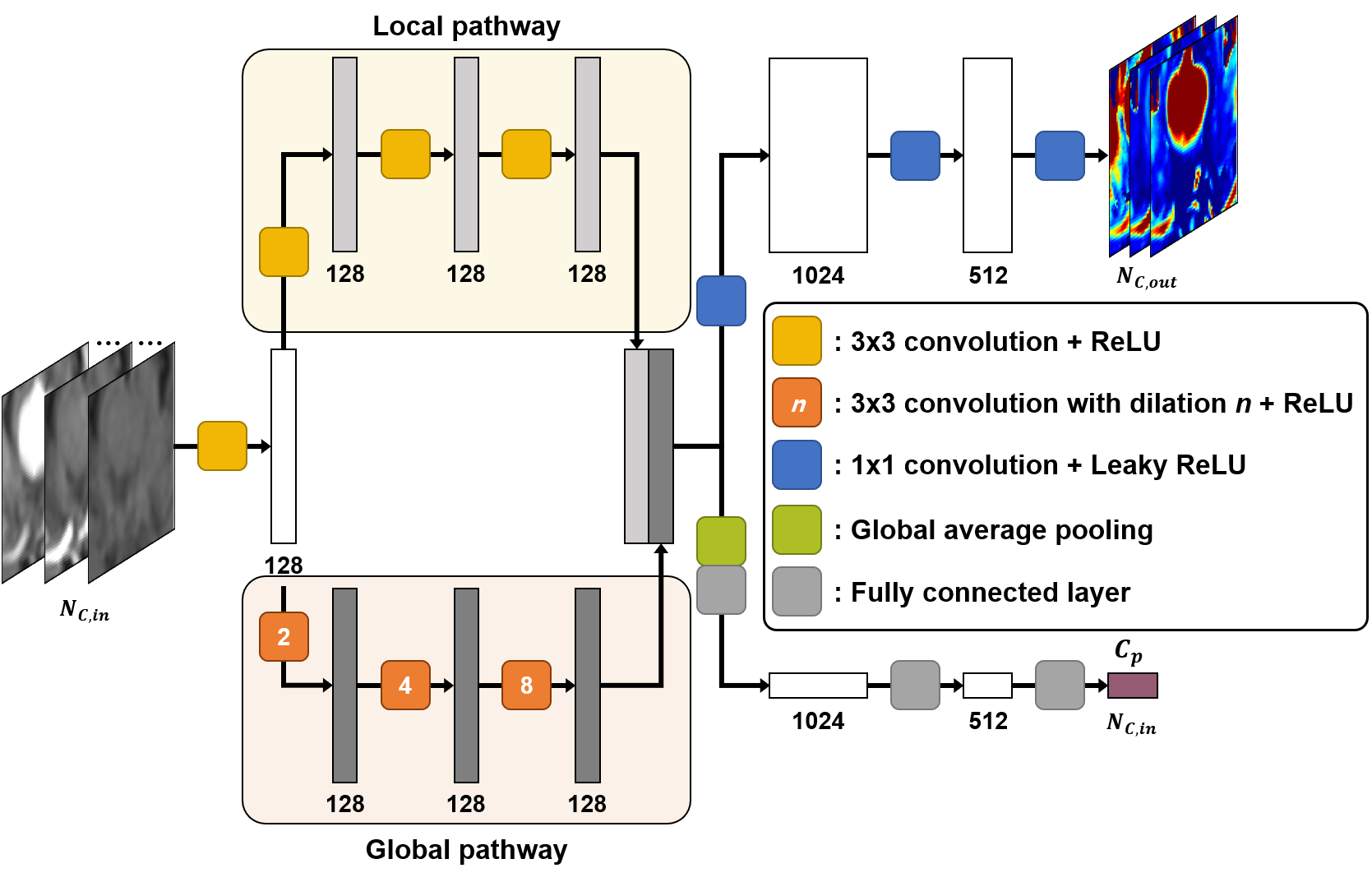}}
\caption{The architecture of the generator comprises two pathways: the local pathway and the global pathway.
The local pathway focuses on capturing local features, while the global pathway aims to capture global context information.
The local and global features are then merged, and the split pathways estimate the PK parameters and $C_p$.
The number below each block indicates the number of channels.}
\label{fig:generator}
\end{figure}

\subsection{Network Architecture}
The CycleGAN architecture for PK parameter estimation is illustrated in Fig. \ref{fig:cycleDCE}.
Unlike the vanilla CycleGAN, which employs two generators and two discriminators, our CycleGAN, designed for the deterministic physics model from PK parameters to DCE-MRI, utilizes only one generator and one discriminator.
To generate DCE-MRI signals from PK parameter maps, the pre-contrast image intensity $S_0$, $T_1$ map, $C_p$, and scan parameters such as repetition time, flip angle, the time interval between frames, and relaxivity of the contrast agent are required.
However, during inference, only the time series DCE-MRI is required as input to the generator, eliminating the need for additional measurements such as $T_1$ or $C_p$.

The generator architecture is depicted in Fig. \ref{fig:generator}.
The input consists of concatenated time frames of DCE-MRI, resulting in a channel dimension equal to the number of time frames $N_{C, \text{in}}$.
The initial convolutional layer extracts feature from the input.
The generator is then divided into two pathways, as seen in previous works \cite{ulas2019convolutional,fang2021convolutional}.
The local pathway employs three $3\times3$ convolutional layers to capture local details, while the global pathway utilizes three dilated convolutional layers with different dilations to capture global information.
The outputs of the local and global pathways are concatenated, and this combined feature is fed into two separate pathways.
The first pathway estimates the PK parameter maps using three $1\times1$ convolutional layers.
The number of channels in the output  $N_{C, \text{out}}$ corresponds to the number of parameters to be estimated, which is 3 for the eTofts model and 2 for the Patlak model.
In the second pathway, $C_p$ is estimated using global average pooling and fully connected layers.
The number of channels in $C_p$ matches the number of input channels ($N_{C, \text{in}}$), which represents the number of time frames.

\begin{figure}[!t]
\centerline{\includegraphics[width=\columnwidth]{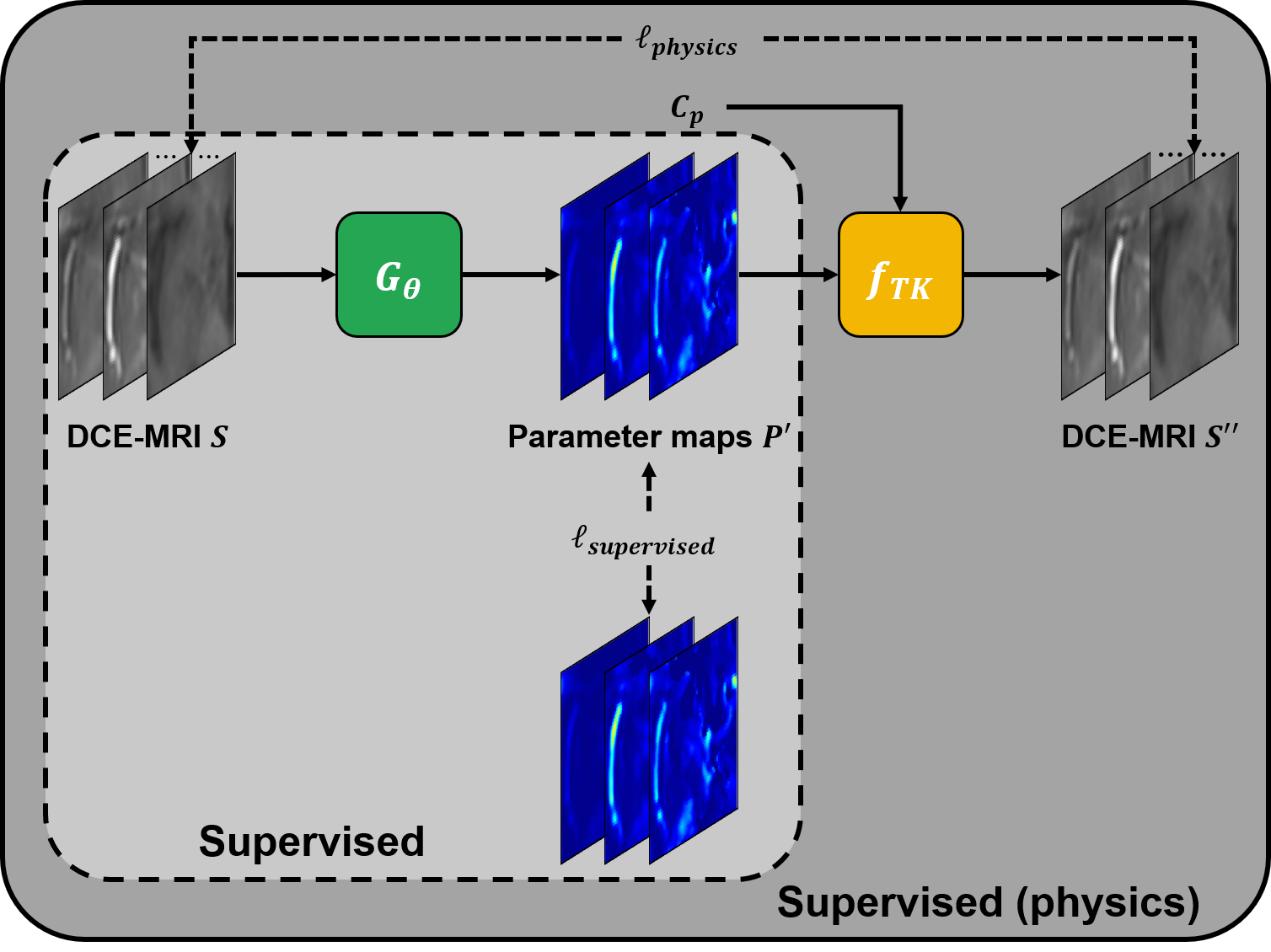}}
\caption{The supervised methods for comparison.
`\textbf{Supervised}' refers to the supervised method trained with L1 loss between the output and label, while `\textbf{Supervised (physics)}' denotes the supervised learning approach incorporating the physics model.}
\label{fig:supervised}
\end{figure}

For the discriminator, we adopt the PatchGAN discriminator introduced in \cite{zhu2017unpaired}.
It consists of $4\times4$ convolutions, instance normalization \cite{ulyanov2016instance}, and leaky ReLU activation.
The initial number of filters in the discriminator is set to 32.

\begin{figure*}[!t]
\centering
\includegraphics[width=0.85\linewidth]{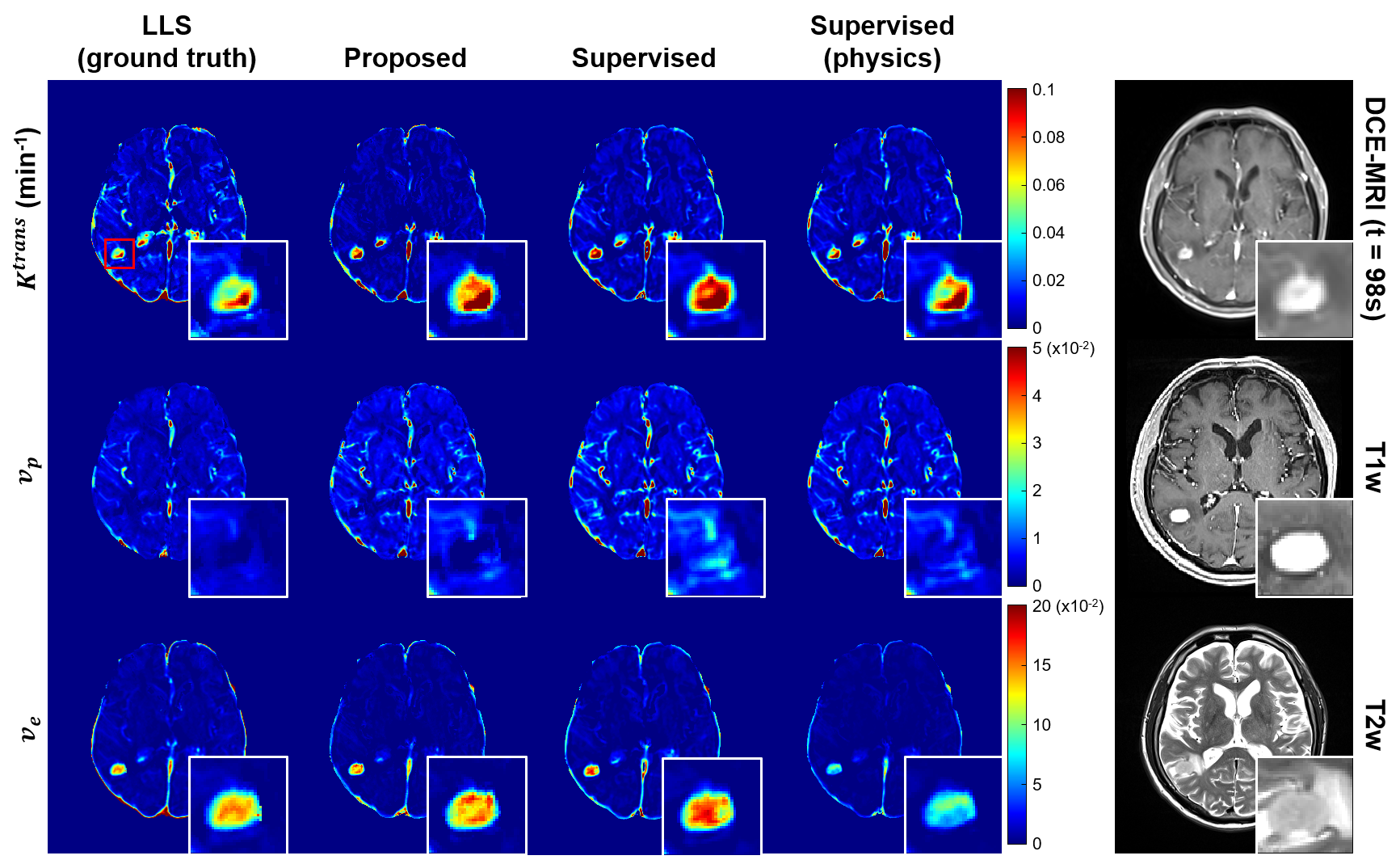}
\caption{Results of PK parameter estimation methods for simulated `Tumor Data' (eTofts model). Here, 
the LLS corresponds to the ground-truth data, from which dynamic images are generated using eTofts model for inference by various models.
The enlarged images highlight the tumor region.
The scale bar for each parameter is located to the right of the parameter maps.
}
\label{fig:tumor_etofts_simul}
\end{figure*}

\renewcommand{\arraystretch}{1.2}
\begin{table}[!t]
\caption{Quantitative evaluation results of various methods using simulated data.}
\label{tab:quantitative_results}
\begin{center}
\resizebox{\columnwidth}{!}{
\begin{tabular}{c | c | c | c  c}
\hline
Data    & PK parameter    & Method    & PSNR (dB) & SSIM \\
\hline\hline
\multirow{9}{*}{\makecell{Tumor Data\\(eTofts)}}              & \multirow{3}{*}{$K^{trans}$} & Proposed              & \textbf{38.45} & 0.968 \\
\cline{3-5}
\                                                             &                              & Supervised            & 37.61          & 0.969  \\
\cline{3-5}
\                                                             &                              & Supervised (physics)  & 37.65          & \textbf{0.969} \\
\cline{2-5}
\                                                             & \multirow{3}{*}{$v_p$}       & Proposed              & \textbf{39.49} & \textbf{0.976} \\
\cline{3-5}
\                                                             &                              & Supervised            & 39.16          & 0.973  \\
\cline{3-5}
\                                                             &                              & Supervised (physics)  & 39.40          & 0.975 \\
\cline{2-5}
\                                                             & \multirow{3}{*}{$v_e$}       & Proposed              & \textbf{43.41} & \textbf{0.986} \\
\cline{3-5}
\                                                             &                              & Supervised            & 43.30          & 0.985  \\
\cline{3-5}
\                                                             &                              & Supervised (physics)  & 43.35          & 0.986 \\
\hline
\multirow{6}{*}{\makecell{MCI Data\\(Patlak)}}                & \multirow{3}{*}{$K^{trans}$} & Proposed              & \textbf{40.50} & 0.977 \\
\cline{3-5}
\                                                             &                              & Supervised            & 40.12          & 0.976  \\
\cline{3-5}
\                                                             &                              & Supervised (physics)  & 40.25          & \textbf{0.978} \\
\cline{2-5}
\                                                             & \multirow{3}{*}{$v_p$}       & Proposed              & \textbf{37.96} & \textbf{0.958} \\
\cline{3-5}
\                                                             &                              & Supervised            & 37.62          & 0.952  \\
\cline{3-5}
\                                                             &                              & Supervised (physics)  & 37.61          & 0.951 \\
\hline
\end{tabular}
}
\end{center}
\end{table}
\renewcommand{\arraystretch}{1}

\subsection{Implementation Details}
To handle PK parameters that may exhibit values outside the expected range in extra-brain regions, we employed the brain extraction tool \cite{smith2002fast} to extract brain masks, using only PK parameters within the brain regions.
During training, small patches were cropped from DCE-MRI images or PK parameter maps.
The patch size was set to 48$\times$48, and for every training step, patches were randomly cropped from various locations.
To augment the data, random flipping was applied.
Additionally, the PK parameters were scaled by appropriate factors to match their range.
Specifically, $K^{trans}$, $v_p$, and $v_e$ in the eTofts model were scaled by 20, 40, and 4, respectively, while $K^{trans}$ and $v_p$ in the Patlak model were scaled by 40 and 8, respectively.
During inference, the entire DCE-MRI images were fed into the network.

For the regularization parameters, we chose $\gamma=10$ (Eq. \eqref{eq:total}), and $\rho=1$ for the `Tumor Data' and $\rho=0.1$ for the `MCI Data (Eq. \eqref{eq:cycle_pk}).
We utilized the Adam optimizer with $\beta_1=0.5$, $\beta_2=0.999$, and a batch size of 32 to train our networks.
The initial learning rate was set to $10^{-5}$ with linear decay applied.
Our CycleGAN was trained for 200 epochs and implemented in Python using PyTorch.

\subsection{Comparison Methods}
To validate the effectiveness of the proposed methods, we utilized two types of supervised learning approaches.
Firstly, we employed a supervised method trained with a simple L1 loss, defined as follows:
\begin{equation}
    \ell(\Theta) = \alpha \ell_{supervised}(\Theta)
\end{equation}
where $\ell_{supervised}$ is the L1 loss between the network output and the ground truth, and $\alpha$ is an appropriate parameter.
We select $\alpha = 10$ to maintain consistency with the cycle-consistency loss in the CycleGAN.
Furthermore, we utilize the same network architecture as the generator in the CycleGAN, along with the other predefined hyperparameters.

In addition, we incorporate the physics model into the supervised learning method, following the approach employed in existing methods for PK parameter estimation \cite{ulas2019convolutional,fang2021convolutional}.
In the supervised method with the physics model, the network output is used to reconstruct a time series of DCE images by incorporating the physics model.
Subsequently, an additional loss, $\ell_{physics}$, is computed, and the supervised learning with the physics model is trained using the following loss function:
\begin{equation}
    \ell(\Theta) = \alpha \ell_{supervised}(\Theta) + \beta \ell_{physics}(\Theta)
\end{equation}
where $\ell_{physics}$ is the L1 loss between the input DCE images and the reconstructed DCE images, and $\beta$ is a hyperparameter that controls the weighting $\ell_{physics}$.
We set $\beta = 10$, and use the same settings as the supervised learning method without the physics model.
The overall flow of the supervised methods is illustrated in Fig. \ref{fig:supervised}.


\renewcommand{\arraystretch}{2}
\begin{table}[!t]
\caption{Average PK parameter values in tumor regions of simulated `Tumor Data'.}
\label{tab:quantitative_tumor}
\begin{center}
\resizebox{\columnwidth}{!}{
\begin{tabular}{c | c | c | c}
\hline
PK parameter    & Method    & \makecell{Average\\Value} & \makecell{Different from\\($P < 0.05$)} \\
\hline\hline
\multirow{4}{*}{\makecell{$K^{trans}$\\$(\text{min}^{-1})$}}  & LLS                   & 0.013 & - \\
\cline{2-4}
\                                                             & Proposed              & 0.014 & \makecell{Supervised,\\Supervised (physics)} \\
\cline{2-4}
\                                                             & Supervised            & 0.020 & Proposed  \\
\cline{2-4}
\                                                             & Supervised (physics)  & 0.018 & Proposed \\
\hline
\multirow{4}{*}{\makecell{$v_p$\\$(\times 10^{-2})$}}         & LLS                   & 0.454 & Supervised \\
\cline{2-4}
\                                                             & Proposed              & 0.533 & Supervised \\
\cline{2-4}
\                                                             & Supervised            & 0.575 & \makecell{LLS, Proposed,\\Supervised (physics)}  \\
\cline{2-4}
\                                                             & Supervised (physics)  & 0.440 & Supervised \\
\hline
\multirow{4}{*}{\makecell{$v_e$\\$(\times 10^{-2})$}}         & LLS                   & 4.237 & \makecell{Proposed,\\Supervised (physics)} \\
\cline{2-4}
\                                                             & Proposed              & 1.781 & \makecell{LLS,\\Supervised} \\
\cline{2-4}
\                                                             & Supervised            & 3.403 & \makecell{Proposed,\\Supervised (physics)}  \\
\cline{2-4}
\                                                             & Supervised (physics)  & 2.281 & \makecell{LLS,\\Supervised} \\
\hline
\end{tabular}
}
\end{center}
\end{table}
\renewcommand{\arraystretch}{1}

\section{Experimental Results}\label{sec:result}
\subsection{Simulation Studies}
Although the PK parameter maps obtained from the LLS method are used for training the methods, there are cases where these parameters may not be reliable, especially when the arterial input function (AIF) or other measured parameters are inaccurate.
Therefore, a direct quantitative comparison between the LLS method and deep learning methods may not be appropriate, as it does not account for the potential inaccuracies in the LLS method.
Even if deep learning methods better capture the physics model and yield improved results, the quantitative evaluation alone may not reflect their actual performance.

To address this limitation, we generated simulated DCE time series images based on the physics model and the PK parameters obtained from the LLS method.
Since the simulated data is generated based on the physics model, we can consider the PK parameters from the LLS method as ground truth.
Consequently, deep learning methods that take the simulated data as input should yield similar results to the LLS method, as they effectively capture the underlying physics model.
Therefore, we quantitatively evaluated the deep learning methods using the simulated data.
For this evaluation, we employed two commonly used metrics: the peak signal-to-noise ratio (PSNR) and the structural similarity index measure (SSIM).

Fig. \ref{fig:tumor_etofts_simul} presents the results of PK parameter estimation using various methods on simulated `Tumor Data'.
The simulation assumes that the LLS method provides ground truth results.
Hence, the estimated results from deep learning methods are expected to be similar to those of the LLS method.
However, as depicted in Fig. \ref{fig:tumor_etofts_simul}, the supervised methods, both with and without the physics model, exhibit distinct differences in the estimation of $v_p$ and $v_e$ compared to the LLS method.
Notably, the supervised learning method with the physics model yields relatively lower estimates of $v_e$ in the tumor region compared to the LLS method.
In contrast, our proposed method demonstrates similar estimation results to the LLS method.

The obtained quantitative results of the proposed method and supervised methods are presented in TABLE \ref{tab:quantitative_results}.
Notably, supervised learning demonstrates the inferior results across the quantitative metrics.
This observation suggests that conventional supervised learning approaches do not effectively incorporate the underlying physical model.
While the application of the physics model to supervised learning leads to some improvement in the quantitative results, the gains are relatively modest.
Conversely, our CycleGAN method exhibits notably superior quantitative results compared to the supervised methods.
This outcome strongly suggests that our proposed approach successfully estimates PK parameters by explicitly reflecting the physics model.

\renewcommand{\arraystretch}{2}
\begin{table}[!t]
\caption{Average PK parameter values in hippocampus regions of simulated `MCI Data'.}
\label{tab:quantitative_mci}
\begin{center}
\resizebox{\columnwidth}{!}{
\begin{tabular}{c | c | c | c}
\hline
PK parameter    & Method    & \makecell{Average\\Value} & \makecell{Different from\\($P < 0.05$)} \\
\hline\hline
\multirow{4}{*}{\makecell{$K^{trans}$\\$(\text{min}^{-1})$\\$(\times 10^{-2})$}}  & LLS                   & 0.035 & \makecell{Supervised,\\Supervised (physics)} \\
\cline{2-4}
\                                                                                & Proposed              & 0.039 & \makecell{Supervised,\\Supervised (physics)} \\
\cline{2-4}
\                                                                                & Supervised            & 0.046 & \makecell{LLS,\\Proposed}  \\
\cline{2-4}
\                                                                                & Supervised (physics)  & 0.047 & \makecell{LLS,\\Proposed} \\
\hline
\multirow{4}{*}{\makecell{$v_p$\\$(\times 10^{-2})$}}                            & LLS                   & 1.748 & \makecell{Supervised,\\Supervised (physics)} \\
\cline{2-4}
\                                                                                & Proposed              & 2.327 & Supervised (physics) \\
\cline{2-4}
\                                                                                & Supervised            & 2.532 & LLS  \\
\cline{2-4}
\                                                                                & Supervised (physics)  & 2.648 & \makecell{LLS,\\Proposed} \\
\hline
\end{tabular}
}
\end{center}
\end{table}
\renewcommand{\arraystretch}{1}

\begin{figure*}[!t]
\centering
\includegraphics[width=0.85\linewidth]{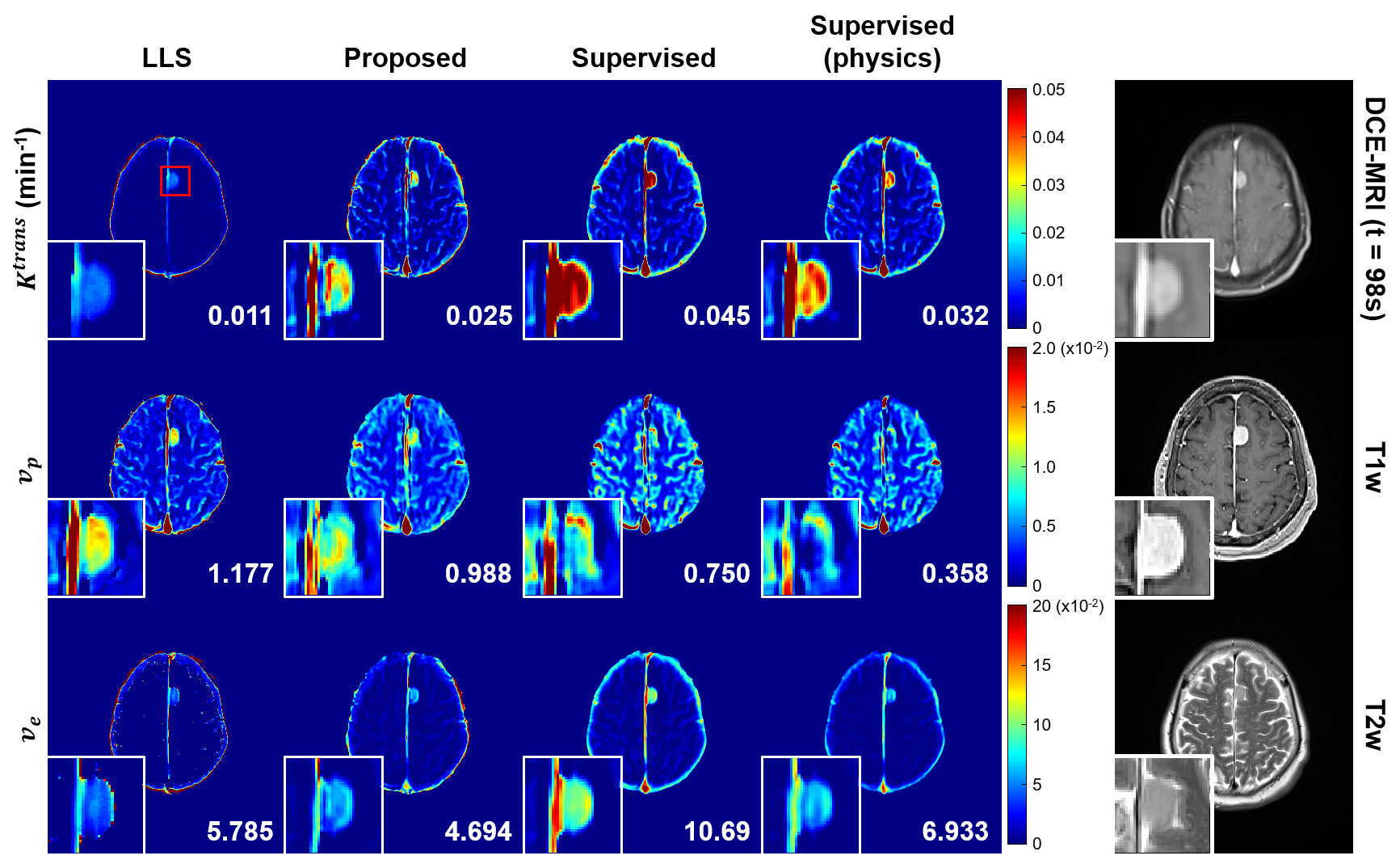}
\caption{Results of PK parameter estimation methods for in vivo `Tumor Data' (eTofts model).
The enlarged images highlight the tumor region.
The number in the bottom right corner indicates the mean parameter values at the tumor region.
The scale bar for each parameter is located to the right of the parameter maps.
}
\label{fig:tumor_etofts}
\end{figure*}

\begin{figure*}[!t]
\centering
\includegraphics[width=0.85\linewidth]{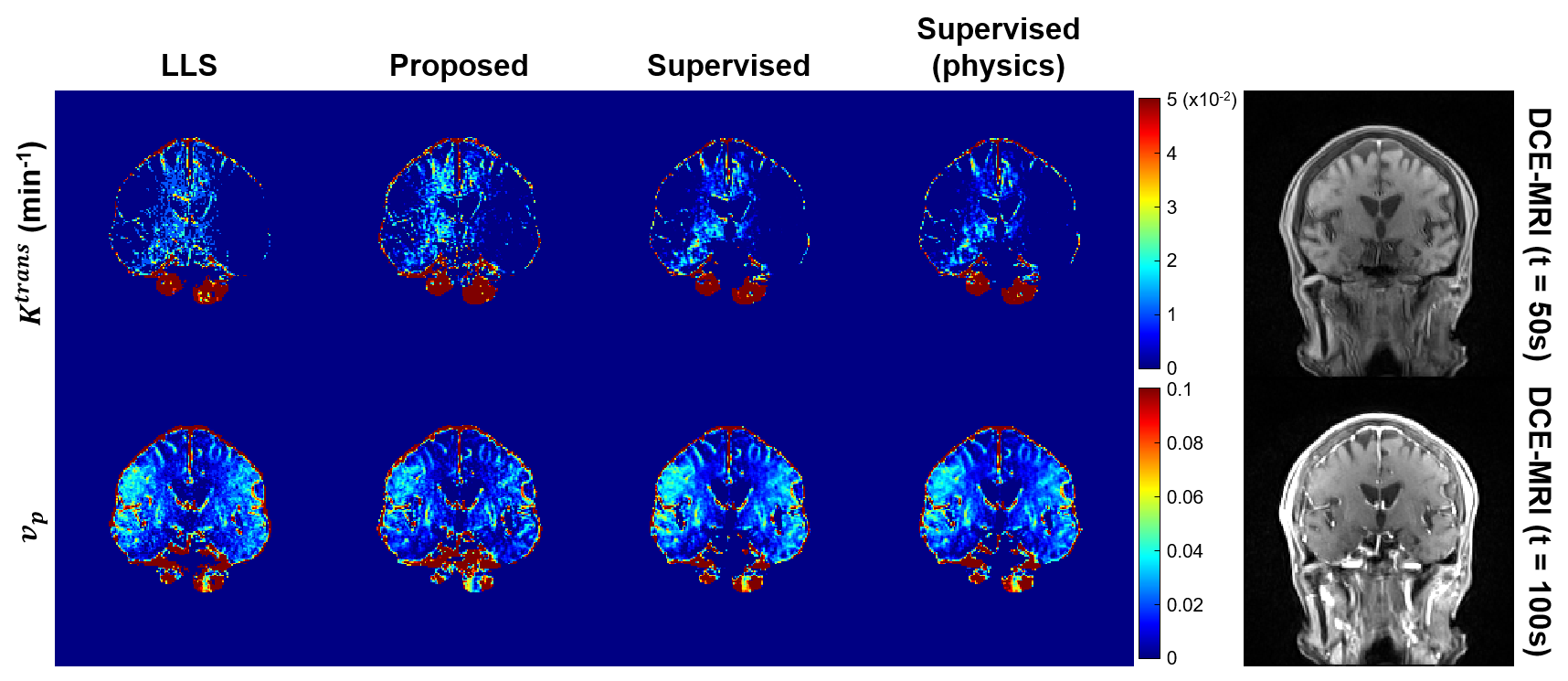}
\caption{Results of PK parameter estimation methods for in vivo `MCI Data' (Patlak model).
The scale bar for each parameter is located to the right of the parameter maps.
}
\label{fig:BB_patlak}
\end{figure*}

We performed a Friedman test to compare the estimated PK parameter values specifically in the tumor regions in simulated `Tumor Data'.
The results of this analysis are presented in TABLE \ref{tab:quantitative_tumor}.
Our proposed method demonstrates average $K^{trans}$ and $v_p$ values that are similar to those estimated by the LLS method in the tumor region.
Conversely, the supervised learning approach tends to exhibit overestimated average values compared to the LLS method.
Additionally, both the proposed method and supervised learning with the physics model show relatively lower average values of $v_e$ compared to the LLS method.
However, it is important to note that these values remain within an acceptable range, as evidenced by the results presented in Fig. \ref{fig:tumor_etofts_simul} and TABLE \ref{tab:quantitative_results}.

TABLE \ref{tab:quantitative_mci} presents the average PK parameter values in the hippocampus region of the simulated `MCI Data'.
Consistently, both the LLS method and our proposed method demonstrate similar average values, indicating that our approach effectively incorporates the underlying physics model during PK parameter estimation.
However, the supervised methods, with or without the physics model, yield results that differ from those obtained by the LLS method.

\subsection{In Vivo Results}
Subsequently, we conducted experiments using in vivo DCE-MRI data.
As the accuracy of the LLS method relies on additional inputs such as AIF, its results may not always be reliable.
Hence, the utilization of in vivo data allows us to verify whether our method can offer more accurate results in comparison to the LLS method.

\paragraph{Qualitative Results}
Fig. \ref{fig:tumor_etofts} presents the estimated PK parameters of the in vivo `Tumor Data' using both the traditional LLS method and deep learning methods.
The enlarged images demonstrate that our proposed CycleGAN method effectively captures the tumor characteristics and produces results that closely resemble those obtained with the LLS method.
However, supervised learning approach yields excessively high $K^{trans}$ and $v_e$ values compared to the LLS method.
When the supervised learning approach is augmented with the physics model, it achieves improved estimation results for $K^{trans}$ and $v_e$ compared to standard supervised learning.
Nonetheless, the supervised learning with the physics model exhibits an abnormally low estimation $v_p$ in the tumor region, in contrast to the $v_p$ estimated by the LLS method or our proposed method.

Additionally, Fig. \ref{fig:BB_patlak} shows the reliable results obtained by our CycleGAN when applied to the Patlak model.
This underscores the versatility and effectiveness of our approach beyond specific model applications.

\begin{figure}[!t]
\centering
\includegraphics[width=0.9\columnwidth]{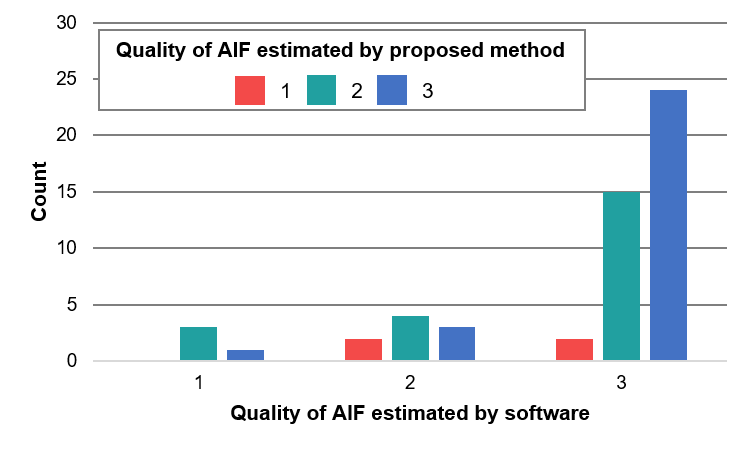}
\caption{AIF quality evaluation results of the software and proposed method using in vivo `Tumor Data' and in vivo `MCI Data'.
}
\label{fig:AIF_evaluation}
\end{figure}

\paragraph{AIF estimation}
The quality of the arterial input function (AIF) is significantly influenced by the data quality or voxel selection.
To mitigate the potential performance degradation caused by AIF quality, our model not only estimates PK parameters but also jointly estimates the AIF.
This integrated approach helps maintain the overall model performance.

To assess the efficacy of our AIF estimation, we conducted a clinical evaluation to evaluate the quality of estimated AIFs.
A radiologist assessed the AIFs generated by NordieICE software and our proposed method using a 3-point scoring system: 1 = poor, 2 = fair, 3 = good.

Fig. \ref{fig:AIF_evaluation} displays the results of the AIF quality evaluation.
The x-axis represents the assessed quality of the AIF obtained from the NordicICE software.
As depicted in Fig. \ref{fig:AIF_evaluation}, our proposed method consistently delivers higher-quality AIF estimations in cases where the AIF derived from the software exhibits poor quality.

\begin{figure}[!t]
\centering
\includegraphics[width=0.7\columnwidth]{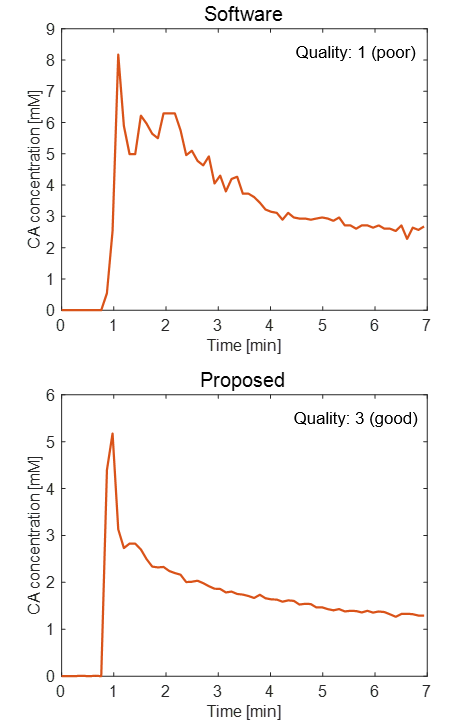}
\caption{Examples of the estimated AIFs generated by the NordicICE software and proposed method.
Both AIFs are from the same in vivo data.
The corresponding quality scores by radiologists are indicated at the top right of each graph.
}
\label{fig:AIF}
\end{figure}

Fig. \ref{fig:AIF} shows a comparison between the AIFs by the software and our proposed method.
It is evident that our method provides more accurate AIF estimations compared to the NordicICE software.

\section{Discussion}\label{sec:discussion}
In summary, we have proposed a novel unpaired deep learning method, based on physics-driven CycleGAN, for pharmacokinetic estimation.
Our findings demonstrate that the proposed CycleGAN method accurately captures tumor characteristics and provides estimation results comparable to the LLS method while reducing computational complexity.
However, supervised learning methods, even when incorporating the physics model, tend to yield exaggerated values for $K^{trans}$ and $v_e$ and exhibit limitations in accurately estimating $v_p$ in the tumor region.
Furthermore, our CycleGAN approach shows consistent and reliable results when applied to the Patlak model.
These results highlight the robustness and applicability of our CycleGAN method for accurate pharmacokinetic parameter estimation.

Additionally, our evaluation of AIF quality demonstrates the superiority of our proposed method.
In cases where the software generates AIFs of poor quality, our approach consistently delivers higher-quality estimations.
This underscores the effectiveness of our method in overcoming limitations associated with suboptimal AIF data.
Moreover, the AIF comparison in Fig. \ref{fig:AIF} provides further support, clearly indicating that our proposed method outperforms the software by providing more accurate AIF estimations.
These results emphasize the value of our approach in obtaining reliable and high-quality AIFs, which contribute to improved pharmacokinetic parameter estimation and enhance the overall performance of the model.

Despite successfully demonstrating accurate PK parameter estimation with reduced runtime and comparable performance to existing methods, our proposed method has certain limitations.
For example, it is important to note that our method is not fully unsupervised and still requires PK parameter maps estimated by NLLS or LLS methods for training, although the training data need not be paired.
Therefore, the performance of our proposed method may be affected if the quality of the training data is not optimal.


Overall, our proposed method presents a promising approach for pharmacokinetic parameter estimation, offering improved accuracy, reduced computational complexity, and the ability to overcome limitations associated with suboptimal AIF data.
Further advancements in training data quality and AIF estimation techniques will contribute to enhancing the performance and applicability of our method.

\section{Conclusion}\label{sec:conclusion}
This paper presents a novel unpaired deep learning method for PK parameter estimation utilizing a physics-driven CycleGAN, based on the physical model of PK parameters.
Our CycleGAN architecture, derived from optimal transport theory, is specifically designed for the known physics model, resulting in a single generator and discriminator.
Despite being trained with unpaired datasets, our method consistently produces reliable PK parameter maps compared to the LLS method and other supervised methods.
Moreover, our proposed method demonstrates superior performance in quantitative evaluations when compared to alternative approaches.
In addition, our proposed method also demonstrates superior performance in estimating the arterial input function, providing more accurate and reliable AIF estimations when the quality of AIFs estimated by existing methods is poor.
We believe that our method can provide valuable insights and accurate PK parameter estimation even in scenarios where paired data is unavailable or missing.

\bibliographystyle{IEEEtran}
\bibliography{ref.bib}

\begin{thebibliography}{10}
\providecommand{\url}[1]{#1}
\csname url@samestyle\endcsname
\providecommand{\newblock}{\relax}
\providecommand{\bibinfo}[2]{#2}
\providecommand{\BIBentrySTDinterwordspacing}{\spaceskip=0pt\relax}
\providecommand{\BIBentryALTinterwordstretchfactor}{4}
\providecommand{\BIBentryALTinterwordspacing}{\spaceskip=\fontdimen2\font plus
\BIBentryALTinterwordstretchfactor\fontdimen3\font minus
  \fontdimen4\font\relax}
\providecommand{\BIBforeignlanguage}[2]{{%
\expandafter\ifx\csname l@#1\endcsname\relax
\typeout{** WARNING: IEEEtran.bst: No hyphenation pattern has been}%
\typeout{** loaded for the language `#1'. Using the pattern for}%
\typeout{** the default language instead.}%
\else
\language=\csname l@#1\endcsname
\fi
#2}}
\providecommand{\BIBdecl}{\relax}
\BIBdecl

\bibitem{verma2012overview}
S.~Verma, B.~Turkbey, N.~Muradyan, A.~Rajesh, F.~Cornud, M.~A. Haider, P.~L.
  Choyke, and M.~Harisinghani, ``Overview of dynamic contrast-enhanced {MRI} in
  prostate cancer diagnosis and management,'' \emph{American Journal of
  Roentgenology}, vol. 198, no.~6, pp. 1277--1288, 2012.

\bibitem{armitage2011use}
P.~A. Armitage, A.~J. Farrall, T.~K. Carpenter, F.~N. Doubal, and J.~M.
  Wardlaw, ``Use of dynamic contrast-enhanced {MRI} to measure subtle
  blood--brain barrier abnormalities,'' \emph{Magnetic resonance imaging},
  vol.~29, no.~3, pp. 305--314, 2011.

\bibitem{van2016blood}
H.~J. Van De~Haar, S.~Burgmans, J.~F. Jansen, M.~J. Van~Osch, M.~A. Van~Buchem,
  M.~Muller, P.~A. Hofman, F.~R. Verhey, and W.~H. Backes, ``Blood-brain
  barrier leakage in patients with early alzheimer disease,'' \emph{Radiology},
  vol. 281, no.~2, pp. 527--535, 2016.

\bibitem{sourbron2013classic}
S.~P. Sourbron and D.~L. Buckley, ``Classic models for dynamic
  contrast-enhanced mri,'' \emph{NMR in Biomedicine}, vol.~26, no.~8, pp.
  1004--1027, 2013.

\bibitem{tofts1999estimating}
P.~S. Tofts, G.~Brix, D.~L. Buckley, J.~L. Evelhoch, E.~Henderson, M.~V. Knopp,
  H.~B. Larsson, T.-Y. Lee, N.~A. Mayr, G.~J. Parker \emph{et~al.},
  ``Estimating kinetic parameters from dynamic contrast-enhanced t1-weighted
  {MRI} of a diffusable tracer: standardized quantities and symbols,''
  \emph{Journal of Magnetic Resonance Imaging: An Official Journal of the
  International Society for Magnetic Resonance in Medicine}, vol.~10, no.~3,
  pp. 223--232, 1999.

\bibitem{patlak1983graphical}
C.~S. Patlak, R.~G. Blasberg, and J.~D. Fenstermacher, ``Graphical evaluation
  of blood-to-brain transfer constants from multiple-time uptake data,''
  \emph{Journal of Cerebral Blood Flow \& Metabolism}, vol.~3, no.~1, pp. 1--7,
  1983.

\bibitem{branch1999subspace}
M.~A. Branch, T.~F. Coleman, and Y.~Li, ``A subspace, interior, and conjugate
  gradient method for large-scale bound-constrained minimization problems,''
  \emph{SIAM Journal on Scientific Computing}, vol.~21, no.~1, pp. 1--23, 1999.

\bibitem{flouri2016fitting}
D.~Flouri, D.~Lesnic, and S.~P. Sourbron, ``Fitting the two-compartment model
  in {DCE-MRI} by linear inversion,'' \emph{Magnetic resonance in medicine},
  vol.~76, no.~3, pp. 998--1006, 2016.

\bibitem{choi2020improving}
K.~S. Choi, S.-H. You, Y.~Han, J.~C. Ye, B.~Jeong, and S.~H. Choi, ``Improving
  the reliability of pharmacokinetic parameters at dynamic contrast-enhanced
  {MRI} in astrocytomas: A deep learning approach,'' \emph{Radiology}, vol.
  297, no.~1, pp. 178--188, 2020.

\bibitem{ulas2019convolutional}
C.~Ulas, D.~Das, M.~J. Thrippleton, M.~d.~C. Valdes~Hernandez, P.~A. Armitage,
  S.~D. Makin, J.~M. Wardlaw, and B.~H. Menze, ``Convolutional neural networks
  for direct inference of pharmacokinetic parameters: Application to stroke
  dynamic contrast-enhanced {MRI},'' \emph{Frontiers in neurology}, vol.~9, p.
  1147, 2019.

\bibitem{fang2021convolutional}
K.~Fang, Z.~Wang, Z.~Li, B.~Wang, G.~Han, Z.~Cheng, Z.~Chen, C.~Lan, Y.~Zhang,
  P.~Zhao \emph{et~al.}, ``Convolutional neural network for accelerating the
  computation of the extended tofts model in dynamic contrast-enhanced magnetic
  resonance imaging,'' \emph{Journal of Magnetic Resonance Imaging}, vol.~53,
  no.~6, pp. 1898--1910, 2021.

\bibitem{sim2020optimal}
B.~Sim, G.~Oh, J.~Kim, C.~Jung, and J.~C. Ye, ``Optimal transport driven
  cyclegan for unsupervised learning in inverse problems,'' \emph{SIAM Journal
  on Imaging Sciences}, vol.~13, no.~4, pp. 2281--2306, 2020.

\bibitem{heye2016tracer}
A.~K. Heye, M.~J. Thrippleton, P.~A. Armitage, M.~d. C.~V. Hern{\'a}ndez, S.~D.
  Makin, A.~Glatz, E.~Sakka, and J.~M. Wardlaw, ``Tracer kinetic modelling for
  dce-mri quantification of subtle blood--brain barrier permeability,''
  \emph{Neuroimage}, vol. 125, pp. 446--455, 2016.

\bibitem{di2009accuracy}
P.~Di~Giovanni, C.~Azlan, T.~S. Ahearn, S.~Semple, F.~J. Gilbert, and T.~W.
  Redpath, ``The accuracy of pharmacokinetic parameter measurement in dce-mri
  of the breast at 3 t,'' \emph{Physics in Medicine \& Biology}, vol.~55,
  no.~1, p. 121, 2009.

\bibitem{chao2017tumourmetrics}
S.-L. Chao, T.~Metens, and M.~Lemort, ``Tumourmetrics: a comprehensive clinical
  solution for the standardization of dce-mri analysis in research and routine
  use,'' \emph{Quantitative imaging in medicine and surgery}, vol.~7, no.~5, p.
  496, 2017.

\bibitem{lim2020cyclegan}
S.~Lim, H.~Park, S.-E. Lee, S.~Chang, B.~Sim, and J.~C. Ye, ``Cyclegan with a
  blur kernel for deconvolution microscopy: Optimal transport geometry,''
  \emph{IEEE Transactions on Computational Imaging}, vol.~6, pp. 1127--1138,
  2020.

\bibitem{villani2008optimal}
C.~Villani, \emph{Optimal transport: old and new}.\hskip 1em plus 0.5em minus
  0.4em\relax Springer Science \& Business Media, 2008, vol. 338.

\bibitem{martin2017wasserstein}
S.~Martin~Arjovsky and L.~Bottou, ``Wasserstein generative adversarial
  networks,'' in \emph{Proceedings of the 34 th International Conference on
  Machine Learning, Sydney, Australia}, 2017.

\bibitem{mao2017least}
X.~Mao, Q.~Li, H.~Xie, R.~Y. Lau, Z.~Wang, and S.~Paul~Smolley, ``Least squares
  generative adversarial networks,'' in \emph{Proceedings of the IEEE
  international conference on computer vision}, 2017, pp. 2794--2802.

\bibitem{szomolanyi2019comparison}
P.~Szomolanyi, M.~Rohrer, T.~Frenzel, I.~M. Noebauer-Huhmann, G.~Jost,
  J.~Endrikat, S.~Trattnig, and H.~Pietsch, ``Comparison of the relaxivities of
  macrocyclic gadolinium-based contrast agents in human plasma at 1.5, 3, and 7
  t, and blood at 3 t,'' \emph{Investigative radiology}, vol.~54, no.~9, p.
  559, 2019.

\bibitem{padhani2002dynamic}
A.~R. Padhani, ``Dynamic contrast-enhanced mri in clinical oncology: current
  status and future directions,'' \emph{Journal of Magnetic Resonance Imaging:
  An Official Journal of the International Society for Magnetic Resonance in
  Medicine}, vol.~16, no.~4, pp. 407--422, 2002.

\bibitem{thrippleton2019quantifying}
M.~J. Thrippleton, W.~H. Backes, S.~Sourbron, M.~Ingrisch, M.~J. van Osch,
  M.~Dichgans, F.~Fazekas, S.~Ropele, R.~Frayne, R.~J. van Oostenbrugge
  \emph{et~al.}, ``Quantifying blood-brain barrier leakage in small vessel
  disease: review and consensus recommendations,'' \emph{Alzheimer's \&
  Dementia}, vol.~15, no.~6, pp. 840--858, 2019.

\bibitem{zhu2017unpaired}
J.-Y. Zhu, T.~Park, P.~Isola, and A.~A. Efros, ``Unpaired image-to-image
  translation using cycle-consistent adversarial networks,'' in
  \emph{Proceedings of the IEEE international conference on computer vision},
  2017, pp. 2223--2232.

\bibitem{ulyanov2016instance}
D.~Ulyanov, A.~Vedaldi, and V.~Lempitsky, ``Instance normalization: The missing
  ingredient for fast stylization,'' \emph{arXiv preprint arXiv:1607.08022},
  2016.

\bibitem{smith2002fast}
S.~M. Smith, ``Fast robust automated brain extraction,'' \emph{Human brain
  mapping}, vol.~17, no.~3, pp. 143--155, 2002.

\end{thebibliography}

\end{document}